\documentclass[aps, twocolumn,superscriptaddress,showpacs,showkeys,amsmath,amssymb,floatfix,nofootinbib]{revtex4}%preprint <-> twocolumn
\usepackage{booktabs,graphicx,mathrsfs,verbatim}
\usepackage[colorlinks=true,citecolor=blue,filecolor=blue,linkcolor=blue,urlcolor=blue,pdftex]{hyperref}
\usepackage[usenames,dvipsnames]{color}
\usepackage{ulem}
\usepackage{fancyhdr}
\usepackage{datetime}

\newcommand{\n}[1]{\mathrm{#1}} % normal (roman) text in math mode

\usepackage[usenames,dvipsnames]{color}

\begin{document}
\title{Comment on ``On the subtleties of searching for dark matter with liquid xenon detectors"}

\newcommand{\note}[1]{\textcolor{red}{#1}}

\newcommand{\assergi}{\affiliation{INFN, Laboratori Nazionali del Gran Sasso, Assergi, 67100, Italy}}
\newcommand{\bologna}{\affiliation{University of Bologna and INFN-Bologna, Bologna, Italy}}
\newcommand{\columbia}{\affiliation{Physics Department, Columbia University, New York, NY 10027, USA}}
\newcommand{\coimbra}{\affiliation{Department of Physics, University of Coimbra, R. Larga, 3004-516, Coimbra, Portugal}}
\newcommand{\heidelberg}{\affiliation{Max-Planck-Institut f\"ur Kernphysik, Saupfercheckweg 1, 69117 Heidelberg, Germany}}
\newcommand{\houston}{\affiliation{Department of Physics and Astronomy, Rice University, Houston, TX 77005 - 1892, USA}}
\newcommand{\laquila}{\affiliation{Department of Physics, University of L'Aquila, 67010, Italy}}
\newcommand{\losangeles}{\affiliation{Physics \& Astronomy Department, University of California, Los Angeles, USA}}
\newcommand{\mainz}{\affiliation{Institut f\"ur Physik, Johannes Gutenberg Universit\"at Mainz, 55099 Mainz, Germany}}
\newcommand{\munster}{\affiliation{Institut f\"ur Kernphysik, Wilhelms-Universit\"at M\"unster, 48149 M\"unster, Germany}}
\newcommand{\nikhef}{\affiliation{Nikhef  and the University of Amsterdam, Science Park, Amsterdam, Netherlands}}
\newcommand{\purdue}{\affiliation{Department of Physics, Purdue University, West Lafayette, IN 47907, USA}}
\newcommand{\shanghai}{\affiliation{Department of Physics, Shanghai Jiao Tong University, Shanghai, 200240, China}}
\newcommand{\subatech}{\affiliation{SUBATECH, Ecole des Mines de Nantes, CNRS/In2p3, Universit\'e de Nantes, 44307 Nantes, France}}
\newcommand{\torino}{\affiliation{University of Torino and INFN-Torino, Torino, Italy}}
\newcommand{\weizmann}{\affiliation{Department of Particle Physics and Astrophysics, Weizmann Institute of Science, 76100 Rehovot, Israel}}
\newcommand{\zurich}{\affiliation{Physics Institute, University of Z\"{u}rich, Winterthurerstr. 190, CH-8057, Switzerland}}

\author{E.~Aprile}\columbia 
\author{M.~Alfonsi}\nikhef
\author{K.~Arisaka}\losangeles
\author{F.~Arneodo}\assergi
\author{C.~Balan}\coimbra
\author{L.~Baudis}\zurich
%\email{laura.baudis@physik.uzh.ch} 
\author{B.~Bauermeister}\mainz
\author{A.~Behrens}\zurich
\author{P.~Beltrame}\losangeles
\author{K.~Bokeloh}\munster
\author{E.~Brown}\munster
\author{G.~Bruno}\assergi
\author{R.~Budnik}\columbia 
\author{J.~M.~R.~Cardoso}\coimbra
\author{W.-T.~Chen}\subatech
\author{B.~Choi}\columbia
\author{D.~Cline}\losangeles
\author{A.~P.~Colijn}\nikhef
\author{H.~Contreras}\columbia
\author{J.~P.~Cussonneau}\subatech
\author{M.~P.~Decowski}\nikhef
\author{E.~Duchovni}\weizmann
\author{S.~Fattori}\mainz
\author{A.~D.~Ferella}\zurich
\author{W.~Fulgione}\torino
\author{F.~Gao}\shanghai
\author{M.~Garbini}\bologna
\author{C.~Ghag}\losangeles
\author{K.-L.~Giboni}\columbia
\author{L.~W.~Goetzke}\columbia
\author{C.~Grignon}\mainz
\author{E.~Gross}\weizmann
\author{W.~Hampel}\heidelberg
\author{F.~Kaether}\heidelberg
\author{A.~Kish}\zurich
\author{J.~Lamblin}\subatech
\author{H.~Landsman}\weizmann
\author{R.~F.~Lang}\purdue\columbia
\author{M.~Le~Calloch}\subatech
\author{C.~Levy}\munster
\author{K.~E.~Lim}\columbia
\author{Q.~Lin}\shanghai
\author{S.~Lindemann}\heidelberg
\author{M.~Lindner}\heidelberg
\author{J.~A.~M.~Lopes}\coimbra
\author{K.~Lung}\losangeles
\author{T.~Marrod\'an~Undagoitia}\zurich
\author{F.~V.~Massoli}\bologna
\author{A.~J.~Melgarejo~Fernandez}\columbia
\author{Y.~Meng}\losangeles
\author{A.~Molinario}\torino
\author{E.~Nativ}\weizmann
\author{K.~Ni}\shanghai
\author{U.~Oberlack}\mainz\houston
\author{S.~E.~A.~Orrigo}\coimbra
\author{E.~Pantic}\losangeles
\author{R.~Persiani}\bologna
\author{G.~Plante}\columbia
\author{N.~Priel}\weizmann
\author{A.~Rizzo}\columbia
\author{S.~Rosendahl}\munster
\author{J.~M.~F.~dos Santos}\coimbra
\author{G.~Sartorelli}\bologna
\author{J.~Schreiner}\heidelberg
\author{M.~Schumann}\zurich
\author{L.~Scotto~Lavina}\subatech
\author{P.~R.~Scovell}\losangeles
\author{M.~Selvi}\bologna
\author{P.~Shagin}\houston
\author{H.~Simgen}\heidelberg
\author{A.~Teymourian}\losangeles
\author{D.~Thers}\subatech
\author{O.~Vitells}\weizmann
\author{H.~Wang}\losangeles
\author{M.~Weber}\heidelberg
\author{C.~Weinheimer}\munster

\collaboration{The XENON100 Collaboration}\noaffiliation

\begin{abstract}

In a recent manuscript (arXiv:1208.5046) Peter Sorensen claims that XENON100's upper limits on spin-independent WIMP-nucleon cross sections for WIMP masses below 10\,GeV ``may be understated by one order of magnitude or more".  Having performed a similar, though more detailed analysis prior to the submission of our new result (arXiv:1207.5988), we do not confirm these findings. We point out the rationale for not considering the described effect in our final analysis and list several potential problems with his study.
	
\end{abstract}



\maketitle

%\linenumbers

In a recent manuscript~\cite{ref::sorensen}, P. Sorensen examines our results from a 225 live-days dark matter run with XENON100~\cite{aprile:2012nq} and claims that the XENON100 upper limit on WIMP-nucleon cross sections at WIMP masses below 10\,GeV might be significantly stronger than our published result. We are aware of the raised issues and take the opportunity to comment here. While we welcome the author's endorsement of our main conclusion, namely the lack of an observed dark matter signal in this run, we do not support his statement of one order of magnitude improvement in sensitivity for low-mass WIMPs after having performed a similar analysis prior to the submission of our manuscript to PRL.

We agree with the argument that in principle one might use the additional information carried by the proportional light signal, S2, in order to obtain a better measure of the energy of each scattering event in our detector.  We would thus exploit not only S1, the prompt scintillation signal, but the fully available phase space.  Indeed, as shown in \cite{Aprile:2011dd} we have used the combined S1 and S2 information to significantly improve the energy resolution of our detector for interactions of gamma rays at various energies and to understand its main background sources \cite{Aprile:2011vb}. On the other hand, as we discuss in more detail later, we are still unable to use the information in the S2-channel at the energies of interest to a dark matter search, for we lack measurements of the ionization yield, $Q_y$, of liquid xenon for nuclear recoils of a few keV.
We also agree with the statement that low-mass WIMPs are expected to show a different S2/S1 versus S1 distribution than the one expected from calibration data with an  $^{241}$AmBe neutron source. In fact, we have studied these effects in detail, in a similar fashion as followed in the paper by P. Sorensen: we have inferred $Q_y$  based on our  $^{241}$AmBe nuclear recoil calibration data and on the measured $\mathcal{L}_\n{eff}(E_\n{nr})$ and have used Monte Carlo simulations to generate the expected distributions of events in the S2/S1 versus S1 parameter space for various WIMP masses.  We have analyzed the influence on our upper limit on WIMP-nucleon cross sections using the same profile likelihood  method as described in \cite{Aprile:2011hx,aprile:2012nq}. For a WIMP mass of 6\,GeV, the upper limit changes from 8.33$\times$10$^{-40}$cm$^2$ to 6.45$\times$10$^{-40}$cm$^2$, at 90\% confidence level, which means an improvement by a factor of 1.3. The difference becomes smaller with increasing WIMP masses, being a factor of 1.2 for a 50\,GeV WIMP. 

At present, we do not consider this effect in our final analysis reported in \cite{aprile:2012nq} for the following reasons:

\begin{itemize}

\item
No direct measurements of $Q_y$ at the low nuclear recoil energies relevant for the low-mass WIMP region exist, as also shown in Figure 2 (upper panel)  in the P. Sorensen manuscript. We thus depend on its inference based on $\mathcal{L}_{\text{eff}}$ and the measured detector response to an  $^{241}$AmBe neutron calibration source. Such an inference, while {\it a priori} viable, requires the simultaneous disentanglement of the $Q_y$ behaviour at low energies from efficiencies in S1 and S2 and thus introduces additional systematic errors into the analysis. We remark that direct measurements of $Q_y$, using small liquid xenon time projection chambers (TPCs) are in progress within our collaboration. Such an independent measurement of the ionization yield will greatly reduce the systematics involved in this approach and will make it a viable option for future analyses of dark matter data from liquid xenon TPCs.  In contrast to $Q_y$, direct measurements of the light yield, parameterized via  $\mathcal{L}_{\text{eff}}$, exist \cite{Aprile:2008rc}, the lowest measured data point being at 3\,keV nuclear recoil energy  \cite{Plante:2011hw}. Indirect determinations of  $\mathcal{L}_{\text{eff}}$, as attempted in~\cite{Sorensen:2008ec} and \cite{Lebedenko:2008gb}, have not yielded results compatible with direct measurements \cite{Plante:2011hw}.  These difficulties are further high-lighted by a re-analysis of the result in \cite{Lebedenko:2008gb} for the same experiment \cite{Horn:2011wz}, which are now compatible with data from direct $\mathcal{L}_{\text{eff}}$ determinations.

\item
The reason why the  $^{241}$AmBe calibration data do not accurately describe the expected S2/S1 versus S1 distribution of low-mass WIMPs can be mainly ascribed to four factors: i) The threshold in S1 is much higher in energy than the corresponding threshold in S2; ii) The resolution in S1 in our TPC is much coarser than the resolution in S2; iii) The steep energy spectra and kinematic cut-off of low-mass WIMPs are well below the mean energy threshold set by the S1
measurement; iv) S2 signals do not fluctuate upwards coherently with S1, but are rather independent processes.  As a result, an upward fluctuation of a nuclear recoil well below the mean energy threshold in S1 will most likely result in an S2 close to its mean corresponding to the low energy. The corresponding threshold in S2 is much lower, allowing us to sample the bulk of the distribution in S2 but only the tail in S1.  
The response for low-mass WIMPs therefore systematically moves towards the lower left corner of our two-dimensional dataspace.

\item
We have performed a profile likelihood analysis~\cite{Aprile:2011hx} for the case of WIMP-induced interactions distributed qualitatively similar to Figure~4 in \cite{ref::sorensen}. In this study, we have used  a mean $\mathcal{L}_\n{eff}(E_\n{nr})$ as shown in~\cite{aprile:2012nq} and a $Q_y$ derived from comparing our $^{241}$AmBe calibration data to a dedicated Monte Carlo study which includes detailed knowledge of our detector and event selection cut efficiencies \cite{ref::ambe}. As stated above, at the lowest WIMP mass of 6\,GeV at which we show an exclusion limit in~\cite{aprile:2012nq}, our result would only improve by a factor of 1.3,  the improvement becoming less and less significant for higher WIMP masses. The reason why the limits improve is this: as the expected WIMP distribution moves downwards in $\log_{10}$(S2/S1), its separation from the electronic recoil background distribution increases, which in turn enhances the signal-to-background discrimination. The effect is fully exploited in the profile likelihood analysis, which does not use a pre-defined discrimination cut in the S2/S1 versus S1 space. Nonetheless, such an analysis would not increase the sensitivity of the experiment significantly, while introducing additional systematic errors.  We have thus decided to publish a conservative result using a better known, S1-based energy scale and to define the WIMP signal region based on $^{241}$AmBe neutron calibration data down to the lowest WIMP masses that can be probed by our experiment.

\end{itemize}

We add four additional comments to highlight the problems with the analysis performed by P. Sorensen: 

\begin{itemize}

\item
The overall cut acceptance, as shown in Figure 1 in \cite{aprile:2012nq}, is not constant at 85\%, as assumed in \cite{ref::sorensen}, but varies with energy, decreasing to $\sim$50\% at S1=3 photoelectrons. In addition, the acceptance of the S2$>$150 photoelectrons condition is set to zero below S1=1 photoelectron, as shown in the same figure and detailed in \cite{Aprile:2012vw}. This is effectively equivalent to not considering events below 3\,keV nuclear recoil energy, where no ${\cal L}_\n{eff}$ measurements exist. Ignoring these cuts, as apparently done by P. Sorensen, will lead to an overestimated sensitivity increase, in particular at low WIMP masses. 

\item
The  ``detector-specific details'' used in \cite{ref::sorensen} were obtained from \cite{Aprile:2010um}, which describes our detector during its commissioning run in 2009. The science data presented in \cite{aprile:2012nq} were acquired under different conditions, which will affect the analysis. For instance, the electron lifetime, determined by the liquid xenon purity,  was considerably higher during the last run, an effect which has a direct impact on the size of the detected S2 signals. Moreover, we use only the S2 signals detected by the bottom PMT array for computing the S2/S1-ratio \cite{aprile:2012nq}, and the electron extraction field was slightly lower in this run.

\item
For our result reported in~\cite{aprile:2012nq}, the uncertainty in ${\cal L}_\n{eff}$ is taken into account in the profile likelihood approach and a limit with 90\% C.L. coverage is derived. This has to be included in the analysis presented in~\cite{ref::sorensen}, otherwise  the result is too optimistic. In addition, it is not clear if and how the known background expectation and its uncertainty for the 225 live-days run is taken into account. 

\item
We agree on the importance of understanding the population with low S2/S1-values below the nuclear recoil region. A detailed study of these events, which are also present in our gamma calibration data of the electronic recoil band, is in progress. We nonetheless remark that the two observed events in XENON100's 225 live-days run  are located within the bulk of the nuclear recoil distribution, and could thus be nuclear recoil candidates. At the same time, unlike the statement in  \cite{ref::sorensen},  these events are also consistent with ``anomalous leakage", as defined in \cite{aprile:2012nq} .

\end{itemize}

\noindent
To summarize, while we agree with P. Sorensen's considerations on the expected signal region for low-mass WIMPs \cite{ref::sorensen}, we have observed only a mild impact on our analysis results, in contrast to the claimed ``one order of magnitude or more" improvement in sensitivity. At present, we refrain from using this approach, for a robust analysis should be based on measurements rather than indirectly inferred quantities.  Once direct measurements of the charge yield of  nuclear recoils at low energies will become available, we will undoubtedly make use of this additional information to redefine our energy scale and strengthen our analysis potential for low-mass WIMPs.


\begin{thebibliography}{26}

 \bibitem{ref::sorensen}
 P.~Sorensen, arXiv:1208.5046v1 [astro-ph.CO].
 
\bibitem{aprile:2012nq} 
  E. Aprile, {\it et al.}  [XENON100 Collaboration],
  %``Dark Matter Results from 225 live-days of XENON100 Data,''
  arXiv:1207.5988 [astro-ph.CO].
  %%CITATION = ARXIV:1207.5988;%%
  
  %\cite{Aprile:2011dd}
\bibitem{Aprile:2011dd} 
  E.~Aprile {\it et al.}  [XENON100 Collaboration],
  %``The XENON100 Dark Matter Experiment,''
  Astropart.\ Phys.\  {\bf 35}, 573 (2012)
  arXiv:1107.2155 [astro-ph.IM].
  %%CITATION = ARXIV:1107.2155;%%
  
  %\cite{Aprile:2011vb}
\bibitem{Aprile:2011vb} 
  E.~Aprile, K.~Arisaka, F.~Arneodo, A.~Askin, L.~Baudis, A.~Behrens, K.~Bokeloh and E.~Brown {\it et al.},
  %``Study of the electromagnetic background in the XENON100 experiment,''
  Phys.\ Rev.\ D {\bf 83}, 082001 (2011)
  [Erratum-ibid.\ D {\bf 85}, 029904 (2012)]
  arXiv:1101.3866 [astro-ph.IM].
  %%CITATION = ARXIV:1101.3866;%%
  
%\cite{Aprile:2011hx}
\bibitem{Aprile:2011hx} 
  E.~Aprile {\it et al.}  [XENON100 Collaboration],
  %``Likelihood Approach to the First Dark Matter Results from XENON100,''
  Phys.\ Rev.\ D {\bf 84}, 052003 (2011)
  arXiv:1103.0303 [hep-ex].
  %%CITATION = ARXIV:1103.0303;%%

%\cite{Aprile:2008rc}
\bibitem{Aprile:2008rc} 
  E.~Aprile, L.~Baudis, B.~Choi, K.~L.~Giboni, K.~Lim, A.~Manalaysay, M.~E.~Monzani and G.~Plante {\it et al.},
  %``New Measurement of the Relative Scintillation Efficiency of Xenon Nuclear Recoils Below 10 keV,''
  Phys.\ Rev.\ C {\bf 79}, 045807 (2009)
  arXiv:0810.0274 [astro-ph].
  %%CITATION = ARXIV:0810.0274;%%
  
 \bibitem{Plante:2011hw} 
  G.~Plante, E.~Aprile, R.~Budnik, B.~Choi, K.~L.~Giboni, L.~W.~Goetzke, R.~F.~Lang and K.~E.~Lim {\it et al.},
  %``New Measurement of the Scintillation Efficiency of Low-Energy Nuclear Recoils in Liquid Xenon,''
  Phys.\ Rev.\ C {\bf 84}, 045805 (2011)
  arXiv:1104.2587 [nucl-ex].
  %%CITATION = ARXIV:1104.2587;%%
  
%\cite{Sorensen:2008ec}
\bibitem{Sorensen:2008ec} 
  P.~Sorensen, A.~Manzur, C.~E.~Dahl, J.~Angle, E.~Aprile, F.~Arneodo, L.~Baudis and A.~Bernstein {\it et al.},
  %``The scintillation and ionization yield of liquid xenon for nuclear recoils,''
  Nucl.\ Instrum.\ Meth.\ A {\bf 601}, 339 (2009)
  arXiv:0807.0459 [astro-ph].
  %%CITATION = ARXIV:0807.0459;%%
   
%\cite{Lebedenko:2008gb}
\bibitem{Lebedenko:2008gb} 
  V.~N.~Lebedenko, H.~M.~Araujo, E.~J.~Barnes, A.~Bewick, R.~Cashmore, V.~Chepel, A.~Currie and D.~Davidge {\it et al.},
  %``Result from the First Science Run of the ZEPLIN-III Dark Matter Search Experiment,''
  Phys.\ Rev.\ D {\bf 80}, 052010 (2009)
  arXiv:0812.1150 [astro-ph].
  %%CITATION = ARXIV:0812.1150;%%
  
%\cite{Horn:2011wz}
\bibitem{Horn:2011wz} 
  M.~Horn, V.~A.~Belov, D.~Y.~.Akimov, H.~M.~Araujo, E.~J.~Barnes, A.~A.~Burenkov, V.~Chepel and A.~Currie {\it et al.},
  %``Nuclear recoil scintillation and ionisation yields in liquid xenon from ZEPLIN-III data,''
  Phys.\ Lett.\ B {\bf 705}, 471 (2011)
  [arXiv:1106.0694 [physics.ins-det]].
  %%CITATION = ARXIV:1106.0694;%%
  
\bibitem{ref::ambe}
 E.~Aprile {\it et al.}  [XENON100 Collaboration], to be published.

 %\cite{Aprile:2012vw}
\bibitem{Aprile:2012vw} 
  E.~Aprile {\it et al.}  [XENON100 Collaboration],
  %``Analysis of the XENON100 Dark Matter Search Data,''
  arXiv:1207.3458 [astro-ph.IM].
  %%CITATION = ARXIV:1207.3458;%% 

%\cite{Aprile:2010um}
\bibitem{Aprile:2010um} 
  E.~Aprile {\it et al.}  [XENON100 Collaboration],
  %``First Dark Matter Results from the XENON100 Experiment,''
  Phys.\ Rev.\ Lett.\  {\bf 105}, 131302 (2010)
  arXiv:1005.0380 [astro-ph.CO].
  %%CITATION = ARXIV:1005.0380;%%
 

\end{thebibliography}
\end{document}